\documentstyle[aps,floats,psfig,twocolumn]{revtex} 
\def\gz{\ifmmode{Z\hskip -4.8pt Z}
    \else{\hbox{$Z\hskip -4.8pt Z$}}\fi} 

\newcommand{\be}{\begin{equation}}
\newcommand{\ee}{\end{equation}}
\newcommand{\bea}{\begin{eqnarray}}
\newcommand{\eea}{\end{eqnarray}}

\begin{document}
\tighten
\draft
\title{Spin Excitations in La$\mbox{\boldmath$_2$}$CuO$\mbox{\boldmath$_4$}$ : Consistent Description by Inclusion of Ring-Exchange}

\author{A.~A. Katanin$^{a,b}$ and A.~P. Kampf $^a$} 
\address{
$^a$ Institut f\"ur Physik, Theoretische Physik III, 
Elektronische Korrelationen und Magnetismus,\\
Universit\"at Augsburg, 86135 Augsburg, Germany\\
$^b$ Institute of Metal Physics, 620219 Ekaterinburg, Russia}
\address{~
\parbox{14cm}{\rm 
\medskip
We consider the square lattice Heisenberg antiferromagnet with plaquette 
ring exchange and a finite interlayer coupling leading to a consistent 
description of the spin-wave excitation spectrum in La$_2$CuO$_4$. The values 
of the in-plane exchange parameters, including ring-exchange J$_\Box$, are 
obtained consistently by an accurate fit to the experimentally observed 
in-plane spin-wave dispersion, while the out-of-plane exchange interaction is 
found from the temperature dependence of the sublattice magnetization at low
temperatures. The fitted exchange interactions $J=151.9$meV and 
$J_{\Box}=0.24J$ give values for the spin stiffness and the N\'eel 
temperature in excellent agreement with the experimental data.
}}      

\maketitle

The magnetic properties of La$_2$CuO$_4$ have been the subject of many 
detailed investigations over the last decade. Understanding this undoped parent
compound of high temperature superconducting cuprates is a precondition for 
the many theories which describe metallic cuprates by doping carriers into a 
layered antiferromagnet.
The conventional starting point for undoped cuprates is the two-dimensional 
(2D) spin-1/2 Heisenberg model with only the nearest-neighbor exchange 
interaction $J$ \cite{Manousakis}, which thereby provides the important 
magnetic energy scale needed as input to theories for the metallic and 
superconducting properties of doped cuprates. Despite the 
substantial progress on the theory of the 2D Heisenberg antiferromagnet 
\cite{Chakravarty}, which includes such physical properties as the temperature 
dependence of the magnetic correlation length \cite{Hasenfratz,Carretta}, some 
of the experimental facts for La$_2$CuO$_4$ have clearly demonstrated that a 
complete description of the magnetic excitations requires additional physics 
not contained in the 2D Heisenberg model with $J$ only. Examples include the 
asymmetric lineshape of the two-magnon Raman intensity \cite{Lyons}
or the infrared optical absorption \cite{Perkins}, which have led to proposals 
that spin-phonon interactions \cite{Nori}, resonant phenomena 
\cite{ChubukovFrenkel,Schonfeld}, purely fermionic contributions \cite{Ho}, or 
cyclic ring-exchange \cite{Honda,Eroles,Lorenzana} need to be included. 
In particular, the importance 
of ring (plaquette) exchange has very recently found direct experimental 
support from the observed dispersion of the spin-waves along the magnetic 
Brillouin zone boundary \cite{Aeppli}.

The 4-spin plaquette ring-exchange interaction $J_\Box$ was considered rather 
early as a possible non-negligible correction to the nearest-neighbor 
Heisenberg model \cite{Roger89,Schmidt}. This higher-order spin coupling 
arises naturally in a strong coupling $t/U$ expansion to fourth order for the 
single band Hubbard model at half-filling \cite{Takahashi77,MacDonald90}. Its 
quantitative significance was recently demonstrated in high order perturbation 
expansions \cite{Band3} and {\it ab initio} cluster calculations in realistic 
three-band Hubbard
models for the CuO$_2$ planes \cite{Calzado}. These derivations of effective 
spin models for the low-energy magnetic properties of the undoped CuO$_2$ 
planes led to the estimate $J_\Box/J=0.11$. A linear spin-wave analysis  
of the spectrum in La$_2$CuO$_4$ at 10K in Ref. \cite{Aeppli} has deduced a 
considerably larger value $J_\Box/J=0.41$. The necessity for a sizeable 
$J_\Box$ has recently also been conjectured for the spin ladder compound 
La$_2$Ca$_8$Cu$_{24}$O$_{41}$\cite{Matsuda}.

The spin-wave theory and the quasiclassical phase diagram of the frustrated 
Heisenberg model with ring exchange was investigated in Ref. \cite{Chubukov}.
However, the quantum and thermal renormalizations have not so far been 
taken into account in the spin-wave theory. It is well known for the
quasi-2D Heisenberg model without the ring-exchange term (see Ref.\cite{SSWT} 
and references therein) that such renormalizations can
substantially change the excitation spectrum of the system. The authors of
Ref. \cite{Aeppli} considered the simplest renormalization of the spectrum by
allowing for an overall quantum renormalization factor, which was obtained 
for the 2D Heisenberg model with nearest-neighbor exchange within a $1/S$ 
expansion to order $1/S^2$ \cite{LargeS} and by series expansion from the 
Ising limit \cite{Singh95}. However, in the absence
of a consistent analysis of the spin-wave renormalization in the presence 
of ring-exchange or a next-nearest neighbor coupling, it is not possible to 
provide an accurate determination of exchange parameters. Indeed, as we show 
below, the effects of quantum and thermal fluctuations are not simply captured 
by a single renormalization factor, and a consistent treatment of the spin-wave
spectrum with $J_\Box\neq$0 to order $1/S$ reveals that the previous early 
estimate $J$=136meV from high energy neutron scattering \cite{Hayden} or 
two-magnon Raman scattering \cite{SinghFleury} requires 
a correction at least as large as 10\%. Also, we show that the recent estimate 
$J_\Box=0.41J$ in Ref. \cite{Aeppli} appears to be twice as large as the value 
calculated by accounting systematically for 1/S renormalizations.

In this paper we consider the corrections to the spin-wave
spectrum to first order in $1/S$ for finite $J_{\Box }$ using a
self-consistent spin-wave theory \cite{SSWT}. We obtain values of the 
in-plane and interplane exchange interactions of La$_2$CuO$_4$ allowing 
an accurate fit of the dispersion. We verify that the obtained exchange 
interactions correctly reproduce the measured values for the spin stiffness 
and the N\'eel temperature.

We start from the Heisenberg model with ring-exchange 
\cite{Roger89,Takahashi77,MacDonald90} 
\begin{eqnarray}
H &=&{J\over 2}\sum_{i,\delta }{\bf S}_i\cdot {\bf S}_{i+\delta }+{J^\prime
\over 2}\sum_{i,\delta ^{\prime }}{\bf S}_i\cdot {\bf S}_{i+\delta^{\prime}}+
{J^{\prime\prime}\over 2}\sum_{i,\delta ^{\prime\prime }}{\bf S}_i\cdot 
{\bf S}_{i+\delta ^{\prime \prime }}  \nonumber \\
&+& {J_{\perp}\over 2}\sum_{i,\delta _{\perp }}{\bf S}_i\cdot {\bf S}%
_{i+\delta _{\perp }}+J_{\Box }\sum_{\langle ijkl\rangle }\big[ ({\bf S}%
_i\cdot {\bf S}_j)({\bf S}_k\cdot {\bf S}_l)  \label{H} \\
&+&({\bf S}_i\cdot {\bf S}_l)({\bf S}_k\cdot {\bf S}_j)-({\bf S}%
_i\cdot {\bf S}_k)({\bf S}_j\cdot {\bf S}_l)\big]  \nonumber
\end{eqnarray}
where $J$, $J^{\prime}$, and $J^{\prime \prime }$ are the first ($\delta $),
second ($\delta ^{\prime }$) and third 
($\delta ^{\prime \prime }$)-nearest-neighbor in-plane exchanges, 
$\delta_\perp$ connects to the nearest-neighbor sites in the adjacent planes 
with interplane exchange $J_\perp$, and
$\langle ijkl\rangle$ denotes the four sites of a planar plaquette involved in
the ring exchange.
We use the Dyson-Maleev representation for the spin operators 
\begin{equation}
\left. 
\begin{array}{l}
S_i^{+}=\sqrt{2S}a_i\,,\;S_i^z=S-a_i^{\dagger }a_i \\ 
S_i^{-}=\sqrt{2S}(a_i^{\dagger }-\displaystyle{1\over 2S}a_i^{\dagger }
a_i^{\dagger }a_i)
\end{array}
\right\} \,i\in A\,\,,  \label{BKJa}
\end{equation}
\begin{equation}
\left. 
\begin{array}{l}
S_j^{+}=\sqrt{2S}b_j^{\dagger }\,,\;S_j^z=-S+b_j^{\dagger }b_j \\ 
S_j^{-}=\sqrt{2S}(b_j-\displaystyle{1\over 2S}b_j^{\dagger }b_jb_j)
\end{array}
\right\} \,j\in B\,\,  \label{BKJb}
\end{equation}
where $A$ and $B$ denote the magnetic sublattices; $a_i^{\dagger },a_i,$ and 
$b_j^{\dagger },b_j$ are Bose operators. After substituting Eqs. (\ref{BKJa}%
) and (\ref{BKJb}) into the Hamiltonian (\ref{H}), we decouple quartic terms
into quadratic ones according to the procedure described in Ref. \cite{SSWT}%
. Keeping consistently all terms to order $1/S$, we obtain 
\begin{eqnarray}
H &=&S\sum_i\sum_{d=\delta ,\delta _{\perp }}J_d\gamma _d(D_i^{\dagger
}D_i+D_i^{\dagger }D_{i+d})  \nonumber \\
&&-S\sum_i\sum_{d=\delta ^{\prime },\delta ^{\prime \prime }}J_d\gamma
_d(D_i^{\dagger }D_i-D_i^{\dagger }D_{i+d}) \\
&&-J_{\Box }S^3\sum_i\sum_{d=\delta ,\delta ^{\prime }}(\gamma _0^{\Box
}D_i^{\dagger }D_i-\gamma _d^{\Box }D_i^{\dagger }D_{i+d})  \nonumber
\end{eqnarray}
where $D_i=a_i$ for $\,i\in A\,$ and $D_i=b_i^{\dagger }$ for $\,i\in B;\;$%
also we use the notation $J_\delta =J,\;J_{\delta ^{\prime }}=J^{\prime }$
etc. The renormalization of the bare exchange parameters due to quantum-
and thermal fluctuations is described by the coefficients:
\begin{eqnarray}
\gamma _d&=&\left\{ 
\begin{array}{cc}
1-[\langle a_i^{\dagger }a_i\rangle +\langle a_ib_{i+d}\rangle ]/S, & 
d=\delta ,\delta _{\perp } \\ 
1-[\langle a_i^{\dagger }a_i\rangle -\langle a_i^{\dagger }a_{i+d}\rangle
]/S, & d=\delta ^{\prime },\delta ^{\prime \prime }
\end{array}
\right.  \, , \label{g1}\\
\;\gamma _0^{\Box } &=&1-[3\langle a_i^{\dagger }a_i\rangle +6\langle
a_ib_{i+\delta }\rangle +3\langle a_i^{\dagger }a_{i+\delta ^{\prime
}}\rangle ]/S\, ,  \nonumber \\
\gamma _\delta ^{\Box } &=&1-[3\langle a_i^{\dagger }a_i\rangle +5\langle
a_ib_{i+\delta }\rangle +2\langle a_i^{\dagger }a_{i+\delta ^{\prime
}}\rangle ]/S\, ,  \nonumber \\
\gamma _{\delta ^{\prime }}^{\Box } &=&1-[3\langle a_i^{\dagger }a_i\rangle
+4\langle a_ib_{i+\delta }\rangle +\langle a_i^{\dagger }a_{i+\delta
^{\prime }}\rangle ]/S\, .  \label{g2}
\end{eqnarray}
Diagonalization of this Hamiltonian yields the spin-wave spectrum 
\begin{eqnarray}
E_{{\bf k}} &=&\sqrt{A_{{\bf k}}^2-B_{{\bf k}}^2}\, ,  \label{Ek} \\
A_{{\bf k}} &=&4S[J\gamma -J^{\prime }\gamma ^{\prime }(1-\nu _{{\bf k}%
}^{\delta ^{\prime }})-J_{\Box }S^2(\gamma _0^{\Box }+\gamma _{\delta
^{\prime }}^{\Box }\nu _{{\bf k}}^{\delta ^{\prime }})]  \nonumber \\
&&\ \ \ \ \ \ \ \ -4J^{\prime \prime }S\gamma ^{\prime \prime }(1-\nu _{{\bf %
k}}^{\delta ^{\prime \prime }})+2J_{\perp }S\gamma _{\perp } \, ,\\
B_{{\bf k}} &=&4S(J\gamma -J_{\Box }S^2\gamma _\delta ^{\Box })\nu _{{\bf k}%
}^\delta +2J_{\perp }S\gamma _{\perp }\nu _{{\bf k}}^{\delta _{\perp }}\, , 
\end{eqnarray}
where 
\begin{eqnarray}
\nu _{{\bf k}}^\delta &=&(\cos k_x+\cos k_y)/2\;,\;\phantom{22}
\nu _{{\bf k}}^{\delta^{\prime}}=\cos k_x\cos k_y\; ,  \nonumber \\
\nu _{{\bf k}}^{\delta ^{\prime \prime }} &=&(\cos 2k_x+\cos 2k_y)/2\;,\;
\nu _{%
{\bf k}}^{\delta _{\perp }}=\cos k_z\, ,
\end{eqnarray}
and $\gamma =\gamma _\delta$, $\gamma^{\prime}=\gamma _{\delta^{\prime }}$
etc.; the lattice constants are set to unity. Since the equality 
$\gamma _0^{\Box }=2\gamma _\delta ^{\Box }-\gamma_{\delta ^{\prime }}^{\Box}$ 
is satisfied, the spectrum given by Eq. (\ref{Ek}) is necessarily 
gapless. It is apparent from this result for the dispersion that the 
renormalization coefficients $\{\gamma_d,\gamma^\Box_d\}$ cannot be combined 
into a single
overall renormalization factor. The averages of the bosonic operators which 
enter in Eqs. (\ref{g1}) and (\ref{g2}) are 
\begin{eqnarray}
\langle a_i^{\dagger }a_i\rangle &=&\sum_{{\bf k}}\frac{A_{{\bf k}}}{2E_{%
{\bf k}}}\coth \frac{E_{{\bf k}}}{2T}-\frac 12  \, ,\label{Av} \\
\langle a_i^{\dagger }a_{i+d}\rangle &=&\sum_{{\bf k}}\frac{A_{{\bf k}}\nu _{%
{\bf k}}^d}{2E_{{\bf k}}}\coth \frac{E_{{\bf k}}}{2T}\, ,  \nonumber \\
\langle a_ib_{i+d}\rangle &=&-\sum_{{\bf k}}\frac{B_{{\bf k}}\nu _{{\bf k}}^d%
}{2E_{{\bf k}}}\coth \frac{E_{{\bf k}}}{2T}\, .  \nonumber
\end{eqnarray}
The expression for the sublattice magnetization reads 
\begin{equation}
\overline{S}=S-\langle a_i^{\dagger }a_i\rangle \, .
\end{equation}

As previously discussed for quasi-2D magnets\cite{SSWT}, Eqs. (\ref{g1}), 
(\ref{g2}), and (\ref{Av}) must be solved self-consistently. The spin-wave 
velocity $c$ is obtained by expanding the dispersion at small 
wavevector $k=\sqrt{k_x^2+k_y^2}$, leading to
\begin{equation}
E_{{\bf k}}\simeq ck\;,\;c=2\sqrt{2}Z_cJS\, ,  \label{c}
\end{equation}
where 
\begin{eqnarray}
Z_c &=&(A_{\bf 0}/4JS)^{1/2}\left[ \gamma -2(J^{\prime }/J)\gamma ^{\prime
}-4(J^{\prime \prime }/J)\gamma ^{\prime \prime }\right.  \nonumber \\
&&\ \ \ \ \left. +2(J_{\Box }/J)S^2(\gamma _{\delta ^{\prime }}^{\Box
}-\gamma _\delta ^{\Box })\right] ^{1/2}
\label{Zc}
\end{eqnarray}
is the spin-wave velocity renormalization factor. For the spin stiffness we
obtain the result 
\begin{eqnarray}
\rho _s &=& JS^2Z_\rho\, ,\label{rs}\\
Z_\rho &=&(4J\overline{S}_0/A_{\bf 0})Z_c^2\, .
\label{Zrho}
\end{eqnarray}
Given $c$ and $\rho _s$, the transverse susceptibility follows immediately as
\begin{equation}
\chi _{\perp } =\rho _s/c^2=\overline{S}_0/(2SA_{\bf 0}) =Z_\chi /(8J)
\label{Zchi}
\end{equation}
where $Z_\chi =Z_\rho /Z_c^2$. For 
$J^{\prime}=J^{\prime\prime}=J_{\Box }=J_{\perp }=0$, {\it i.e.} the 2D 
Heisenberg model with only nearest-neighbor exchange, the self-consistent 
numerical solution of Eqs. (\ref{g1}) and (\ref{Av}) at zero temperature gives 
\cite{Takahashi89,Yoshioka,SSWT} $\gamma=1.158$ and $\overline{S}_0=0.303$, 
which corresponds to $Z_c=1.158,\;Z_\rho =0.702$, and $Z_\chi =0.524$. We note 
that due to the relations (\ref{Zrho}) and (\ref{Zchi}), $Z_\rho$ and $Z_\chi$ 
contain partially contributions of order $1/S^2$. The above numbers are close 
to those found for the 2D nearest-neighbor Heisenberg model in a systematic 
$1/S$ 
expansion to order $1/S^2$: $Z_c=1.179,\;Z_\rho =0.724$, and $Z_\chi =0.514$ 
\cite{LargeS}. However, we show below, the presence of next-nearest 
neighbor and plaquette ring-exchange terms substantially alters these numbers.

We use Eqs. (\ref{g1}), (\ref{g2}), and (\ref{Av}) at zero temperature to fit 
the experimentally determined planar spin-wave dispersion using the data at 
T=10K from
Ref. \cite{Aeppli}. To restrict the number of fitting parameters, we suppose 
$J^{\prime }=J^{\prime \prime }$. This restriction is well justified by the
perturbation expansions of the half-filled one- and three-band Hubbard
models \cite{Takahashi77,MacDonald90,Band3}. The inelastic neutron
scattering data together with our fit result along a selected path in the
Brillouin zone at $T=10$K are shown in Fig. 1. The best fit is obtained for
the following parameter set: 
\begin{equation}
J=151.9\,\text{meV},\,\,J^{\prime }=J^{\prime \prime }=0.025J,\,\,J_{\Box
}=0.24J.  \label{Js}
\end{equation}
While the values of $J^{\prime}$ and $J^{\prime \prime }$ are practically 
indistinguishable from those in Ref.\cite{Aeppli} and $J$ in (\ref{Js}) is 
only 3\% larger, our extracted value of $J_{\Box}$ is 50\% lower.  

\begin{figure}[t!]
\psfig{file=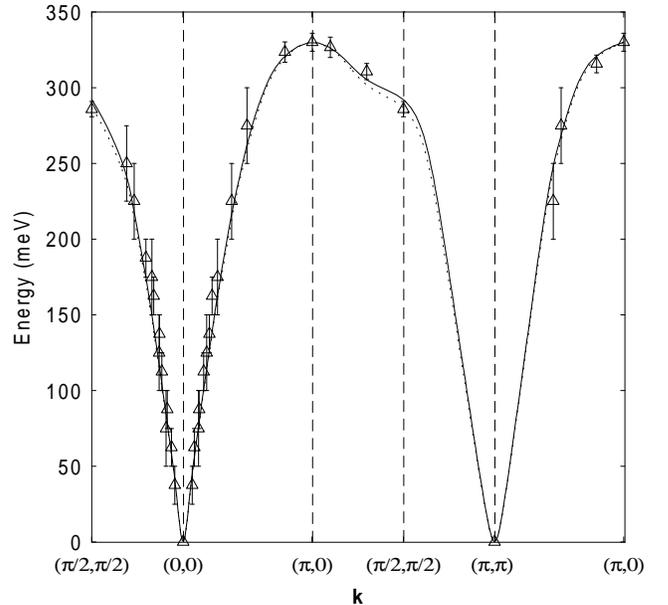,width=90mm,silent=}
\vspace{2mm}
\caption{Spin-wave dispersion along high symmetry directions in the 2D 
Brillouin zone. The triangles are the experimental results of Ref. [14]
for La$_2$CuO$_4$ at 10K. The solid line is the result of a fit to the 
spin-wave dispersion result (7) leading to the exchange couplings as 
listed in Eq. (18). The dashed line is the fit of Ref. [14].}
\label{fig:Dispersion}
\end{figure}

For the corresponding groundstate sublattice magnetization $\overline{S}_0$ 
and for the renormalization parameters $\{\gamma _d,\gamma_d^\Box\}$ we obtain 
\begin{eqnarray}
\overline{S}_0 =0.319,\;\gamma\phantom{\Box}&=&1.158,\;\gamma^{\prime }=0.909,\;\gamma
^{\prime \prime }=0.852\, ,  \label{gn} \\
\gamma _0^{\Box } &=&2.220,\;\gamma _\delta ^{\Box }=1.971,\;\gamma _{\delta
^{\prime }}^{\Box }=1.721\, .  \nonumber
\end{eqnarray}
In our notation, the spectrum used in Ref. \cite{Aeppli} corresponds to
equal renormalization factors: $\gamma =\gamma ^{\prime }=\gamma ^{\prime
\prime }=\gamma _i^{\Box }\simeq 1.18$; the parameter values obtained with
this spectrum were $J^{\prime }/J=J^{\prime \prime }/J=0.020$, and 
$J_{\Box }/J=0.41$. However, some of the $\gamma$-coefficients show a 
remarkable deviation from the value 1.18 found for the 2D system 
with nearest-neighbor exchange only \cite{LargeS}. In particular the 
renormalization coefficients $\{\gamma_d^\Box\}$ for the ring-exchange deviate 
very strongly. We emphasize again, although some fitting parameters are 
very similar to \cite{Aeppli}, it is the self-consitently renormalized 
parameters $\{\gamma_d,\gamma_d^\Box\}$ which allow us to obtain an accurate 
and reliable set of {\it bare} superexchange couplings.

In the self-consistent spin-wave theory presented here the in-plane magnon 
spectrum varies only weakly with temperature at $T\ll J$. Although the 
spectrum changes qualitatively in the same way as found experimentally, it 
accounts only for a few per cent of the observed changes in the zone boundary 
dispersion of the data at $T=295K$ in Ref. \cite{Aeppli}. 

From the parameter values obtained above we deduce the spin stiffness, the 
spin-wave 
velocity, and the transverse susceptibility: $\rho _s=23.8$ meV, $%
c=206$ meV and $\chi _{\perp }=4.8\times 10^{-5}$K$^{-1}.$ The corresponding
values of the renormalization factors as calculated from Eqs. (\ref{Zc}), 
(\ref{Zrho}), and (\ref{Zchi}) are $Z_c=0.96$, $Z_\rho =0.63$ and 
$Z_\chi =0.68$. These values differ substantially from those
for the 2D nearest-neighbor Heisenberg model. We note that our value for the 
spin stiffness is in very good agreement with the earlier estimate
\cite{Keimer} $\rho _s=23.9$meV found from fitting the spin-spin correlation 
length $\xi(T)$ at $T>T_N$ to the non-linear $\sigma$-model result 
\cite{Chakravarty} with the correct preexponential factor\cite{Hasenfratz}.
This agreement strongly supports the validity of the self-consistent 
renormalized spin-wave theory.

As discussed in Refs. \cite{SSWT,Quasi2D}, it is difficult to fit the value 
$J_{\perp }$ from measurements of the out-of-plane spin-wave spectrum. Instead,
we use an alternative procedure and fit the temperature dependence of the 
sublattice magnetization at temperatures $T<T_N/2$ where the above theory is 
reliable ({\it cf}. Ref.\cite{SSWT} and references therein) employing the 
exchange parameters listed in (\ref{Js}). In this way we obtain 
$J_{\perp }/J=1.0\times 10^{-3}$ and $\gamma _{\perp }/\gamma =5.67\times
10^{-4}$ which practically coincides with the previous estimate in Refs. 
\cite{SSWT,Quasi2D}.

As another test of the above results, we also calculate the N\'eel temperature 
for the obtained exchange parameter values. On the basis of a 
renormalization-group approach and a $1/N$ expansion in the $O(N)$ quantum 
nonlinear $\sigma$-model, the result for the N\'eel temperature of a quasi-2D 
isotropic Heisenberg antiferromagnet has the form 
\cite{SSWT,Quasi2D}
\begin{equation}
T_N=4\pi \rho _s\left[ \ln \frac{T_N^2}{c^2\alpha _r}+3\ln \frac{4\pi \rho _s%
}{T_N}-0.0660\right] ^{-1}
\end{equation}
where $\alpha _r=(\gamma _{\perp }/\gamma )_{T=0}$, and $c$ and $\rho_s$ are 
the respective groundstate spin-wave velocity and spin stiffness
given by Eqs. (\ref{rs}) and (\ref{c}). With the above parameter values 
(\ref{Js}) we obtain $T_N=328K$, in almost perfect agreement with the 
experimental value $T_N=325K$ \cite{Keimer}.

In conclusion, we have considered the renormalization of the spin-wave 
spectrum to order $1/S$ for the Heisenberg antiferromagnet in the presence of 
plaquette ring-exchange. The results allow for an accurate fit of the magnon 
dispersion in La$_2$CuO$_4$ and a consistent determination of the exchange 
coupling parameters for this material. As an independent check of the 
parameter set, the spin stiffness and the N\'eel temperature are 
correctly reproduced. With $J=151.9$meV the obtained value for the bare
ring-exchange coupling $J_\Box/J=0.24$ in La$_2$CuO$_4$ is significant 
although due to its 4-spin coupling the ring-exchange term in the Hamiltonian 
will be reduced by a factor $S^2$ with respect to the $J$ term. The
magnitude of $J_\Box$ suggests that for hole doped cuprates ring-exchange 
might be relevant, too, and we may postulate that it is 
connected to recent proposals of staggered circulating currents in underdoped 
materials \cite{Varma,Laughlin}. The role of ring-exchange for the spectral 
lineshape of the $B_{1g}$ shift in Raman experiments and for infrared 
absorption remains to be reexplored and contrasted to the recent proposal of a 
purely fermionic origin of spectral weight at higher energies \cite{Ho}.

It is a pleasure to thank B. Normand and T. Kopp for useful discussions. We 
are indebted to G. Aeppli for sending us experimental data for the spin wave 
dispersion. We acknowledge support through 
Sonderforschungsbereich 484 of the Deutsche Forschungsgemeinschaft.


\begin{references}
\bibitem{Manousakis}  E. Manousakis, Rev. Mod. Phys. {\bf 63}, 1 (1991).

\bibitem{Chakravarty}  S. Chakravarty, B.~I. Halperin, and D.~R. Nelson,
Phys. Rev. B {\bf 39}, 2344 (1989).

\bibitem{Hasenfratz}  P. Hasenfratz and F. Niedermayer, Phys. Lett. B {\bf %
268}, 231 (1991).

\bibitem{Carretta}  P. Carretta, T. Ciabattoni, A. Cuccoli, E. Mognaschi, A.
Rigamonti, V. Tognetti, and P. Verrucchi, Phys. Rev. Lett. {\bf 84}, 366
(2000).

\bibitem{Lyons} K.~B. Lyons, P.~E. Sulewski, P.~A. Fleury, H.~L. Carter, A.~S. 
Cooper, and G.~P. Espinosa, Phys. Rev. B {\bf 39}, 9693 (1989); S. Sugai et 
al., {\it ibid}. {\bf 42}, 1045 (1990).

\bibitem{Perkins} J.~D. Perkins, J.~M. Graybeal, M.~A. Kastner, R.~J. 
Birgeneau, J.~P. Falck, and M. Greven, Phys. Rev. Lett. {\bf 71}, 1621 (1993).

\bibitem{Nori}  F. Nori, R. Merlin, S. Haas, A.~W. Sandvik, and E. Dagotto,
Phys. Rev. Lett. {\bf 75}, 553 (1995).

\bibitem{ChubukovFrenkel}  A. Chubukov and D. Frenkel, Phys. Rev. Lett. {\bf %
74}, 3057 (1995).

\bibitem{Schonfeld}  F. Sch\"onfeld, A.~P. Kampf, and E. M\"uller-Hartmann,
Z. Phys. B {\bf 102}, 25 (1997).

\bibitem{Ho} C.-M. Ho, V.~N. Muthukumar, M. Ogata, and P.~W. Anderson, Phys. 
Rev. Lett. {\bf 86}, 1626 (2001).

\bibitem{Honda} Y. Honda, Y. Kuramoto, and T. Watanabe, Phys. Rev. B {\bf 47},
11329 (1993).

\bibitem{Eroles}  J. Eroles, C.~D. Batista, S.~B. Bacci, and E.~R. Gagliano,
Phys. Rev. B {\bf 59}, 1468 (1999).

\bibitem{Lorenzana}  J. Lorenzana, J. Eroles, and S. Sorella, Phys. Rev.
Lett. {\bf 83}, 5122 (1999).

\bibitem{Aeppli}  R. Coldea, S.~M. Hayden, G. Aeppli, T.~G. Perring, C.~D.
Frost, T.~E. Mason, S.-W. Cheong, and Z. Fisk, Phys. Rev. Lett. {\bf 86},
5377 (2001).

\bibitem{Roger89}  M. Roger and J.~M. Delrieu, Phys. Rev. B {\bf 39}, 2299
(1989).

\bibitem{Schmidt}  H.~J. Schmidt and Y. Kuramoto, Physica (Amsterdam) {\bf %
167C}, 263 (1990).

\bibitem{Takahashi77}  M. Takahashi, J. Phys. C: Solid State Phys. {\bf 10},
1289 (1977).

\bibitem{MacDonald90}  A.~H. MacDonald, S.~M. Girvin, and D. Yoshioka, Phys.
Rev. B {\bf 41}, 2565 (1990); Phys. Rev. B {\bf 37}, 9753 (1988).

\bibitem{Band3}  E. M\"uller-Hartmann and A. Reischl, cond-mat/0105392

\bibitem{Calzado}  C.~J. Calzado and J.-P. Malrieu, cond-mat/0010259

\bibitem{Matsuda}  M. Matsuda, K. Katsumata, R.~S. Eccleston, S. Brehmer,
and H.-J. Mikeska, Phys. Rev. B {\bf 62}, 8903 (2000).

\bibitem{Chubukov}  A. Chubukov, E. Gagliano, and C. Balseiro, Phys. Rev. B 
{\bf 45}, 7889 (1992).

\bibitem{SSWT}  V. Yu. Irkhin, A.~A. Katanin, and M.~I. Katsnelson, Phys.
Lett. A {\bf 157} (1991); Phys. Rev. B {\bf 60}, 1082 (1999).

\bibitem{LargeS}  C.~M. Canali and S.~M. Girvin, Phys. Rev. B {\bf 45}, 7127 
(1992); C.~M. Canali, S.~M. Girvin, and M. Wallin, {\it ibid.} {\bf 45}, 10131 
(1992); I. Igarashi, {\it ibid.} {\bf 46}, 10763 (1992).

\bibitem{Singh95}  R.~R.~P. Singh, Phys. Rev. B{\bf 40}, 7247 (1989);
R.~R.~P. Singh and M.~P. Gelfand, {\it ibid}. {\bf 52}, R15695 (1995).

\bibitem{Hayden}  S.~M. Hayden, G. Aeppli, R. Osborn, A.~D. Taylor, T.~G.
Perring, S.-W. Cheong, and Z. Fisk, Phys. Rev. Lett. {\bf 67}, 3622 (1991).

\bibitem{SinghFleury} R.~R.~P. Singh, P.~A. Fleury, K.~B. Lyons, and P.~E. 
Sulewski, Phys. Rev. Lett. {\bf 62}, 2736 (1989).

\bibitem{Takahashi89}  M. Takahashi, Phys. Rev. B {\bf 40}, 2494 (1989).

\bibitem{Yoshioka}  D. Yoshioka, J. Phys. Soc. Jpn. {\bf 58}, 3733 (1989).

\bibitem{Keimer}  B. Keimer, A. Aharony, A. Auerbach, R.~J. Birgeneau, A. 
Cassanho, Y. Endoh, R.~W. Erwin, M.A. Kastner, and G. Shirane, Phys. Rev. B 
{\bf 45}, 7430 (1992); Phys. Rev. B {\bf 46}, 14034 (1992).

\bibitem{Quasi2D}  V.~Yu. Irkhin and A.~A. Katanin, Phys. Rev. B {\bf 55},
12318 (1997); {\it ibid}. {\bf 57}, 379 (1998).

\bibitem{Varma} C.~M. Varma, Phys. Rev. B {\bf 55}, 14554 (1997), Phys. Rev. 
Lett. {\bf 83}, 3538 (1999).

\bibitem{Laughlin} S. Chakravarty, R.~B. Laughlin, D.~K. Morr, and C. Nayak, 
Phys. Rev. B {\bf 63}, 094503 (2001).
\end{references}
\end{document}